*Pre-print of the manuscript accepted in Acta Materialia*
https://doi.org/10.1016/j.actamat.2019.01.045

# On the structure of defects in the $Fe_7Mo_6$ µ-Phase

S. Schröders[1], S. Sandlöbes[1], B. Berkels[2], S. Korte-Kerzel[1*]

[1]Institut für Metallkunde und Metallphysik, RWTH Aachen University, Kopernikusstraße 14, 52074 Aachen

[2]Aachen Institute for Advanced Study in Computational Engineering Science (AICES), RWTH Aachen University, Schinkelstraße 2, 52062 Aachen

* corresponding author: korte-kerzel@imm.rwth-aachen.de (Sandra Korte-Kerzel)

## Abstract

Topologically close packed phases, among them the µ-phase studied here, are commonly considered as being hard and brittle due to their close packed and complex structure. Nanoindentation enables plastic deformation and therefore investigation of the structure of mobile defects in the µ-phase, which, in contrast to grown-in defects, has not been examined yet. High resolution transmission electron microscopy (HR-TEM) performed on samples deformed by nanoindentation revealed stacking faults which are likely induced by plastic deformation. These defects were compared to theoretically possible stacking faults within the µ-phase building blocks, and in particular Laves phase layers. The experimentally observed stacking faults were found resulting from synchroshear assumed to be associated with deformation in the Laves-phase building blocks.

## 1 Introduction

Amongst the class of topologically close packed phases (TCP), the µ-phase is widely known as an undesirable precipitation in highly alloyed Ni-base superalloys and steels, in particular at high W and Co contents [1-3]. µ-phase precipitates are commonly thought to deteriorate the mechanical properties of these alloys, influencing in particular the creep resistance, yield strength and ductility. The related mechanisms range from pile-up of dislocations which causes microcracking or surface decohesion [4, 5] to grain boundary embrittlement [6, 7] and distortion of unrafted [7] and rafted γ - γ' microstructures [4, 8]. All of these are associated with the hard and brittle nature of TCP phases. However, bent needle-shaped precipitates have also been observed [5, 9], indicating plastic deformation. Nevertheless, to the authors' best knowledge, no attempt has been made to investigate the plastic deformation mechanisms of the µ-phase yet, although the grown-in defect structures have been extensively investigated by high resolution transmission electron microscopy (HR-TEM) [1, 2, 10, 11]. This is mainly due to the intrinsic hardness and brittleness of the µ-phase, originating from the complex and dense packing of TCP phases, which impedes experiments on macroscopic single crystals at low (homologous) temperatures.

The $Fe_7Mo_6$ µ-phase crystallises within the rhombohedral lattice, space group 166, $R\bar{3}m$, with 13 atoms per unit cell [12]. The archetype is $Fe_7W_6$ [12, 13]. In hexagonal notation [13], the parameters of the unit cell given in Figure 1a [14, 15] are **a** = 4.723-4.769 Å and **c** = 25.48-25.83 Å [3, 10, 12]. Figure 1a shows the unit cell with its alternate stacking of irreducible $Zr_4Al_3$ and $MgCu_2$-type layers [1, 16-19]. Faults in this stacking sequence have been observed and reported as stacking faults and growth twins [1, 2, 10, 11, 13], where the $Zr_4Al_3$-layers act as mirror planes [1, 2, 10] with a twinning probability of 0.33 calculated by Carvalho et al. [1]. In addition, non-basal planar faults on $(\bar{1}101)$- and $(1\bar{1}02)$-planes [1, 2, 19] and basal stacking faults with disturbed $Zr_4Al_3$ / $MgCu_2$ stacking order [20] have been reported.

Due to the complex crystal structure of the µ-phase, the direct identification of dislocation paths is challenging. Krämer et al. [21] have suggested that the Peierls barrier is very high in TCP phases due to their close and complex packing ('interlocking'). This is well described for the Laves phases [21-26], the most common group of TCP-phases, which are themselves also composed of a layered structure, i.e. an alternating stacking sequence of single and triple layers. The $MgCu_2$-unit contained in the µ-phase is equivalent to an AαcβB stacking (Figure 1b). The upper and lower single layers, denoted A and B, correspond to Kagomé-nets (Figure 1c) of Fe-atoms and the αcβ sequence is an embedded close packed triple layer (Figure 1b). Within the latter, the smaller atoms of Fe are denoted by Roman letters and the bigger Mo atoms by Greek letters [27]. From this, a full stacking order of αAαcβBβaγCγb…. for a cubic C15 Laves phase results.

Since the µ-phase contains $MgCu_2$-type Laves phase and $Zr_4Al_3$ layers as its building blocks, the deformation mechanism of either or both of these phases might dominate the plasticity of the composite cell of the µ-phase, particularly where it takes place within the stacked layers parallel to the basal plane. While extensive studies have been conducted for Laves phases [21, 23, 26, 28-34], only little has been reported for the $Zr_4Al_3$ phase [35, 36].

The Laves phases are usually brittle at room temperature with a brittle-to-ductile-temperature around 0.6 $T_m$. Above 0.6 $T_m$, plastic deformation by dislocation slip and to a lesser degree by mechanical twinning has been observed [22, 25, 29, 37, 38]. The twinning system is of {111}<112>-type [31] and dislocation slip is based on dislocation with a 1/2<110> Burgers vector in cubic or 1/3<11-20> in hexagonal notation on either (111) or {0001} planes, respectively [21]. Where plastic deformation was achieved at room temperature (RT) or below, it was accommodated mainly by {111}<112> twinning, e.g. for $HfV_2$ + Nb/ Ta, $Cr_2Nb$ and the $MgCu_2$-Laves phases (C15-type) [25, 26, 33, 34, 39]. Using micropillar compression at room temperature, plastic deformation by dislocation slip has been shown for cubic $NbCo_2$ and hexagonal $(Fe,Ni)_2Nb$ Laves phases by Korte et al. [31] and Takata et al. [40].

Based on the stacking sequence of single Kagomé-nets and triple layers, a theoretical model to describe mechanical twinning and dislocation slip in Laves phases was first proposed by Kronberg [41] as synchroshear and later described by Chu, Pope, Hazzledine and Pirouz [28, 42-44]. In synchroshear, two closely-bound partials move through individual neighbouring layers αc and cβ of the αcβ-triple layer (t-layer), as shown schematically in Figure 1d. Movement of the two partials results in a change of the layer's stacking sequence or orientation of close packed atoms viewed along $[11\bar{2}0]$ (Figure 1a) and the resulting layers are denoted t'. In hexagonal Laves phases, both t and t' layers alternate, while the cubic C15 phase and the µ-phase contain only one sequence and the other can form only as growth defect or as a result of synchroshear. Deformation consistent with synchroshear has been reported by Takata et al. [40] and Livingston et al. [39] after low-temperature deformation experiments. First evidence of synchroshear was shown by Allen et al. [27] and Chisholm et al. [45], who imaged lattice faults and a dislocation core by HR-TEM, although it remains unclear if these defects had been induced by plastic deformation.

A substantial body of work therefore exists on the Laves phases, as a constituent of the larger µ-phase unit cell, but the deformation of the latter has hardly been studied. In order to confirm that it is indeed the constituents which determine flow in this larger cell [46] and make the $Fe_7Mo_6$ µ-phase's properties available to researchers concerned with superalloys and steels, a first step involves the investigation of the mobile dislocation structures. Here, this has been accomplished using HR-TEM to characterise faults introduced in nanoindentation at room temperature. Critical stresses and a more statistical assessment of other slip systems may then be obtained in the future using again (nano)indentation and micropillar compression to suppress cracking and consequently allow the analysis of slip traces by SEM and TEM [31, 40, 47-49]. These methods can now be applied also to high temperatures of up to 1000°C in nanoindentation [50], so that ultimately direct relevance to deformation of TCP phases in service may be achieved.

## 2 Experimental details

Ingots of the stoichiometric composition $Fe_7Mo_6$ (45 wt.-% Fe, 55 wt.-% Mo) were prepared by arc-melting and re-melted for five times before homogenization annealing at 800°C for 1000 h. Metallographic sample preparation was done by mechanical grinding and polishing. The samples were examined by optical (Leica DMR, Leica AG) and electron microscopy as well as energy dispersive spectroscopy (EDX) (FEI Helios 600i, FEI Inc.), revealing a multi-phase microstructure of pure µ-phase matrix (55±2 wt.-% Mo) traversed by veins of an iron-rich solid solution and isolated Mo regions, Figure 2. X-ray diffraction (XRD) analysis on a powdered sample revealed diffraction peaks of the µ-phase, as well as of pure iron and molybdenum in descending intensity. No C14 $Fe_2Mo$ Laves phase diffraction peaks were measured. In the XRD spectrum, the main Miller indices of the constituent phases are explicitly shown, additionally, the positions of the diffraction peaks of the µ-phase, α-Fe and α-Mo, are given below the XRD spectrum: blue lines represent the positions of µ-phase peaks, red lines mark the positions of Fe peaks and green lines the positions of Mo peaks. The lattice constants of the µ-phase were refined to **a** = 4.7570(6) Å and **c** = 25.7530(4) Å by employing the Rietveld method [51]. Additionally, electron backscatter diffraction (EBSD), Figure 2, selected area electron diffraction (SAED) analyses and HR(S)-TEM were performed to identify the µ-phase, Figure 3.

Plastic deformation was induced by nanoindentation in µ-phase grains to a maximum load of 200 mN using a Berkovich diamond indenter tip (NanoTest Platform 3, MicroMaterials Ltd.). TEM lamellae were prepared site specifically using a focused ion-beam (FIB) and NanoMill (Model 1040, E.A. Fischione Instruments Inc.). TEM observations were performed on aberration corrected transmission electron microscopes operated at 300 kV (Titan 80-300 TEM and Titan 80-300 STEM, both FEI Inc.). Analysis of the HR-TEM micrographs was performed using the Gatan Microscopy Suite (GMS, Gatan Inc.). Figure 4a shows an atomic force microscope (AFM) image of the indent with the position of the TEM lamella indicated by a dashed rectangle. Figure 4b presents a collage of low resolution TEM micrographs of the TEM lamella highlighting the residual indentation and the area of the high resolution TEM micrographs shown in Figure 6 and 7 located inside the plastic zone of the indent. The orientation of the TEM lamella with respect to the indent is indicated by the centre line A-B of the rectangle in the AFM micrograph.

The HR-(S)TEM images have further been denoised with a variant of the non-local denoising technique "Block-matching and 3D filtering algorithm" (BM3D) that exploits the approximately periodic structure of the depicted crystals [52]. The necessary periodicity information in form of the primitive unit cell was estimated automatically using a real space approach [53].

## 3 Results and Discussion

To differentiate defects induced by deformation from growth defects, characterisation of grown-in defects in the µ-phase was performed, Fig. 5. The defect presented in Fig. 5a is a twin-like stacking fault with the $Zr_4Al_3$-layers as mirror planes, this type of growth defect has been studied and characterised in-depth before [1, 2, 10]. Figure 5b presents a growth twin inducing a 36° rotation of the crystal c-axis, similar to observations by Hiraga et al. [20]. Further, basal stacking faults with disturbed stacking order, specifically a missing $Zr_4Al_3$ layer, were observed using high angle annular dark field (HAADF)-HR-STEM analysis, Fig. 5c. HAADF gives contrast proportional to $Z^p$ (Z is the atomic number and p is an element specific coefficient in the range of 1.4-2), i.e. atoms with a higher Z number appear brighter than atoms with a lower Z number. Consequently, Mo atoms give brighter contrast than Fe atoms in Fig. 5c. The arrows point to the missing $Zr_4Al_3$ layer which is composed of Mo atoms (Fig. 1). No synchroshear configurations were observed in un-deformed material.

Figure 6 shows HR-TEM micrographs and corresponding Fast Fourier Transform (FFT) patterns of defects typically observed below indents in the µ-phase. The defect structure shown in Figure 6a consists of extended basal defects parallel to two of the prismatic planes with a resulting angle of 60° between them. Figure 6b shows a high density of extended basal stacking faults aligned parallel to the same set of prismatic planes. Since these defects are present in the plastic zone below the indent and due to the strain contrast induced by these defects and the crystallographic misorientations across them, we suppose that these defects were formed during deformation. The stacking faults shown in Figure 6b are bound by 60° rotated defects (not shown), similar to those shown in Figure 6a.

$[1\bar{1}00]$ and $[11\bar{2}0]$ zone axes orientations are more suitable than an [0001] zone axis orientation to observe the stacking faults formed by synchroshear. However, several attempts to prepare samples of a suitable thickness and quality with $[1\bar{1}00]$ and $[11\bar{2}0]$ zone axis orientation failed. The samples were either too thick to achieve the resolution needed to characterise such defect structures on the atomic scale or lost their crystallinity when thinned further using a NanoMill® (Fischione). Therefore, the presented more indirect defect analysis using a [0001] zone axis was performed.

Figure 7 shows a higher magnification micrograph of the same type of stacking fault in deformed $Fe_7Mo_6$ µ-phase taken with a B = [0001] zone axis orientation. Bright dots visible in the STEM image represent the positions of ordered columns containing both Mo and Fe atoms in the unit cell. Columns of solely Fe atoms are not resolved due to the lower atomic mass. Planes with a higher density of Mo atoms are visible as lines of higher brightness along the crystallographic planes, corresponding to the Laves layer tetrahedrons. These are well visible in the filtered micrograph (Fig. 7) and highlighted in both, the corresponding insets and the FFT pattern where every third spot along the corresponding $\langle 1\bar{1}00\rangle$ directions, $a_1$ and $a_3$, exhibits higher brightness. The stacking fault is aligned along the $\langle 11\bar{2}0\rangle$ direction. To guide the eye, the crystal unit cell, crystallographic orientation and reference coordinate system are indicated in Figure 7. The stacking fault is bound by another defect with an angle of 60° to the stacking fault, see the lower right corner of the micrographs in Figure 7, here, a slight misorientation across the 60° defect is evident from the slight misalignment of the atomic columns from the [0001] zone axis orientation. The crystal adjacent to the stacking fault has an undistorted appearance. The drawn-in hexagons in the higher resolution micrograph, Figure 7, indicated in red and blue illustrate the crystallographic unit cell and its crystallographic orientation on both sides of the stacking fault. The atomic columns containing Mo and Fe atoms correspond to the crystallographic tetrahedrons of the Laves phase layers. Comparing the alignments of the Laves phase tetrahedrons across the stacking fault, a relative lattice shift associated to a fault of $\frac{1}{3}\langle 1\bar{1}00\rangle$ = 2.74 ± 0.01 Å is visible. The corresponding $\frac{1}{3}\langle 1\bar{1}00\rangle$ vectors are shown in the schematic view of the basal plane.

In order to resolve the origin of this stacking fault, we have analysed stacking fault configurations which can be created by the glide of perfect or partial dislocations, Figure 8, and compared them to the stacking fault observed experimentally in the deformed sample. The images are constructed with a resulting fault along $a_2$, i.e. in the same direction as those imaged in Figure 7. As in the HR-TEM micrographs in Figure 7, the pure Fe columns are not shown and the coloured dots represent Mo / Fe columns in the unit cell with hexagons highlighted by the darker colour.

Figure 8a-d shows the four different stacking fault configurations considered: (a) glide of a partial dislocation with a Burgers vector of $\frac{1}{3}\langle 1\bar{1}00\rangle$ on the basal plane, (b) glide of a perfect dislocation with a Burgers vector of $\frac{1}{3}\langle 11\bar{2}0\rangle$ on the basal plane, (c) glide along $a_2$ by 1/2**a** corresponding to slip along a close-packed direction in the Kagomé-nets, and (d) synchroshear in the triple layer. These scenarios are based on crystallographically possible shear operations which have been reported in the literature [21, 23, 39, 43, 44] for the Laves phases and those feasible in the µ-phase in addition.

The configuration shown in Figure 8a corresponds to single slip by one partial dislocation with the Burgers vector $\frac{1}{3}\langle 1\bar{1}00\rangle$ and causes a spreading of the two halves of the crystal creating a gap; shown for a partial dislocation with Burgers vector $b_2$ creating a gap along $a_2$. The same configuration is created by the glide of single partial dislocations with Burgers vectors $b_1$ and $b_3$ creating a gap along $a_1$ and $a_3$, respectively. This movement of only one partial dislocation with a Burgers vector $\frac{1}{3}\langle 1\bar{1}00\rangle$ therefore appears energetically unfavourable and would require the glide of the corresponding trailing partial, causing a shift by 1**a**. This is a full translation vector of the crystal, and no stacking fault would be created. This configuration corresponds to the glide of a perfect dislocation with a Burgers vector $\frac{1}{3}\langle 11\bar{2}0\rangle$ on the basal plane and is shown in Figure 8b. Indeed, Krämer and Schulze [21] have reported – based on geometrical considerations – that deformation of Laves phases can occur by perfect dislocation glide of dislocations with a Burgers vector of $\frac{1}{3}\langle 11\bar{2}0\rangle$ on basal and prismatic planes. This slip mode has been experimentally confirmed by several groups at temperatures above a homologous temperature of 0.6 $T_m$ for the respective Laves phases [22, 23, 32, 40]. As evident from Figure 8b, glide of a perfect dislocation with a Burgers vector $\frac{1}{3}\langle 11\bar{2}0\rangle$ on the basal plane creates no stacking fault since a full translation of the lattice is created.

Two further proposed dislocation slip modes are not shown as these are similar to those shown in Figure 8a and b: (i) Krämer and Schulze [21] suggested for Laves phases that glide of a partial dislocation with a Burgers vector $\frac{1}{2}\langle 1\bar{1}00\rangle$ may take place on $(11\bar{2}0)$ prismatic planes. In this case, single slip of a $\frac{1}{2}\langle 1\bar{1}00\rangle$ partial dislocation on prismatic planes would create again a "gap" in the crystal without creating a stacking fault of the configuration experimentally measured, while slip of two partial dislocations with a Burgers vector $\frac{1}{2}\langle 1\bar{1}00\rangle$ on prismatic planes results again in a full translation vector of the crystal. (ii) Slip by dislocations with Burgers vectors $\frac{1}{3}\langle 11\bar{2}0\rangle$ and $\frac{1}{3}\langle 1\bar{1}00\rangle$ may also operate in the $Zr_4Al_3$ layers. However, the resulting atomic configurations in the basal plane are equal to those described in detail above for the Laves phase layers.

Figure 8c shows the stacking fault created by the glide of a dislocation with a Burgers vector $\frac{1}{6}\langle 11\bar{2}0\rangle$ on the basal plane. The resulting relative shift across the stacking fault is 1/2**a**, i.e. 2.379 ± 0.004 Å (for the calculation of the error see below), which corresponds neither to the shift measured experimentally in this study (2.74 ± 0.01 Å) nor to values reported for Laves phases before [21, 23, 26, 28-34]. As this vector is not a full translational vector of the crystal, it would create a visible distortion of the shifted crystal, if a lower part of the crystal remains underneath the stacking fault (not shown). This distortion is not apparent in the TEM micrograph.

Lastly, Figure 8d shows the stacking fault configuration created by synchroshear, i.e. the glide of two $\frac{1}{3}\langle 1\bar{1}00\rangle$ synchroshear partial dislocations ($b_2 + b_3$ in this case) in adjacent planes within the triple layer [43, 44].

Glide of the first synchroshear partial produces a stacking fault equal to the one shown in Figure 8a, however, the two synchroshear partial dislocations are closely-bound [43, 44] and glide in conjunction, where the second synchroshear partial dislocation creates a shift "back" of the upper crystal along $b_3$ (Figure 8d) resulting in a shift of $-b_1$. The stacking fault configuration created by the synchroshear mechanism matches the experimentally measured stacking fault configuration (Figure 7).

The construction of this fault is therefore described in more detail and visualised in Figure 9 following the operations identified and described by Hazzledine et al. [43, 44]. Fe atoms are drawn as black circles, the upper layer of Mo atoms of a triple layer as blue circles and those Mo atoms in the lower layer are represented by red circles, these are considered stationary [43, 44]. In Figure 9a the upper row of images gives the atomic configurations seen perpendicular to the (0001)-plane and the lower row of images perpendicular to $(11\bar{2}0)$, along $a_3$. For visibility reasons only one triple layer in the glide plane of the crystal is shown.

The first configuration shown in Figure 9a is the initial one, i.e. the un-deformed perfect crystal. Glide of the first synchroshear partial occurs along $b_2$ (indicated by the red arrow) and produces a stacking fault with the second configuration illustrated in Figure 9a. The glide of this first synchroshear partial moves the Fe layer, and simultaneously the crystal layer above it, so that the Fe layer (labelled c in Figure 1d) shifts onto the places of the upper Mo atoms (β, Figure 1d) in the triple layer. This means that the upper Mo atoms would have to be shifted onto the lower Mo atoms positions (red, α, Figure 1d). However, as these layers are not separated by a large enough interplanar distance, this movement would effectively result in a collision of Mo atoms. The second synchroshear partial therefore glides in sync with the first along $b_3$ to shift the upper Mo atoms into the free spaces left behind by the intermediate layer of Fe atoms, as indicated by the red arrow in configuration 2 of Figure 9a, creating the stacking fault shown in configuration 3 of the same figure. Hence, the resulting net shift of the crystal is the combination of both synchroshear partial Burgers vectors, here $b_2+b_3 = -b_1$, causing a reversal of the orientation of the stacking seen in $[11\bar{2}0]$ direction, i.e. the transformation of a t-(triple)layer into a t'-(triple)layer with a total shift of $-b_1$ [43, 44]. A theoretical value of the shift can be calculated based on the theoretical Burgers vectors and the lattice parameters determined using XRD, where the error is determined from the error in the Rietveld analysis to include a confidence interval of 99% and the correlation of free parameters in powder diffraction [51, 54]. Thus, the theoretical expectation of the shift is given by Equation (*1*):

$$-\,\mathrm{b}_{1,\mathrm{theo}} = \frac{\mathrm{a}}{2\cos 30°} = 2.746 \pm 0.004\ \text{Å} \tag{1}$$

In summary, the stacking fault created by synchroshear is of the same appearance as the experimentally found one (Figure 7) and in excellent agreement with the measured shift from HR-TEM of 2.74 ± 0.01 Å.

Figure 9b represents the stacking fault in the crystal after synchroshear viewed along the crystal c-axis and constructed using the same vectors ($b_2$, $b_3$) as for the stacking fault configuration shown in Figure 8d and Figure 9a. Mo atoms are coloured in red and Fe atoms in black. For better visibility, the Fe atoms in the stacking fault are highlighted in light grey, while the corresponding Mo atoms in the stacking fault are shown in dark blue. Apparently, in the c-axis view, the same atomic configuration as shown in the HR-TEM image in Figure 7 is present. The embedded angle between the stacking fault and the reversed triple layer (t') [45] amounts to 60°, matching the experimentally observed angle shown in Figure 6a and Figure 7.

This indicates that synchroshear mediated plastic deformation in the Laves phase triple layers has induced the stacking faults observed in the vicinity of nanoindents in $Fe_7Mo_6$ µ-phase. As no evidence of slip in the $Zr_4Al_3$ layers could be found here, these results are consistent with those reported in room temperature deformation of pure Laves phase in micropillar experiments by Takata et al. [40], where slip on basal planes was found to occur along the $\langle 1\bar{1}00 \rangle$ direction, i.e. $b_{1,2,3}$. Whether the slight change in lattice parameter of the Laves phase layer in the µ-phase, compared with the pure $Fe_2Mo$ Laves phase, noticeably affects the critical stresses for dislocation motion as seen in some other layered crystals [46], might therefore be investigated in the future using the microcompression technique on both phases.

## 4 Conclusions

We have investigated the post-deformation defect structure of the $Fe_7Mo_6$ µ-phase by means of nanoindentation at room temperature and successive HR-TEM investigation. The µ-phase is characterised by a subcell structure comprised of Laves and $Zr_4Al_3$ layers. Consequently, we have considered stacking fault configurations which can be produced by (partial) dislocation glide in either of these layers and found that the experimentally observed stacking fault is consistent with the one produced by synchroshear in the Laves phase layers. We therefore propose that plastic deformation of the µ-phase parallel to the basal plane is governed by the Laves layers where deformation can occur by synchroshear.

Acknowledgements:

The authors gratefully acknowledge funding by the Deutsche Forschungsgemeinschaft (DFG) in project KO 4603/2-1. Prof. Joachim Mayer is gratefully acknowledged for giving access to the transmission electron microscopes at the Ernst-Ruska-Centre at the Forschungszentrum Jülich. Dr. Hauke Springer and Mr. Michael Kulse at the Max-Planck-Institut für Eisenforschung are acknowledged for synthesis of the $Fe_7Mo_6$ µ-phase. Mr. David Beckers (IMM) is acknowledged for his help in metallographic sample preparation.

Figure Captions:

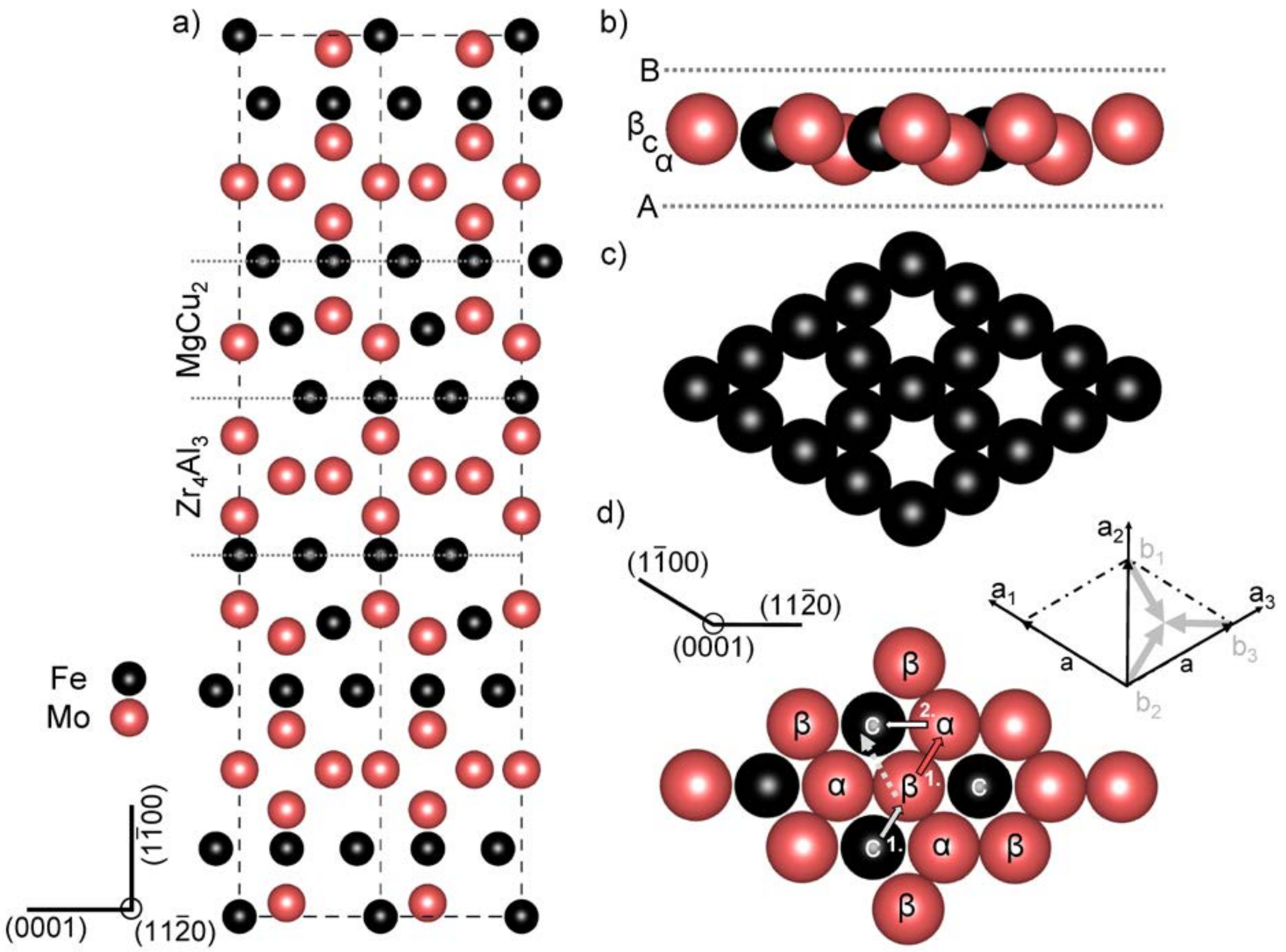

Figure 1: a) Unit cell of the $Fe_7Mo_6$ μ-phase structure visualised [15] perpendicular to $(11\bar{2}0)$ showing the alternating subcell stacking of $Mg_2Cu$-type Laves layers and $Zr_4Al_3$-layers; b) Triple layer of a Laves phase building block indicating the Fe and Mo atom positions between two Fe atom Kagomé-nets A and B, same viewing direction as (a); c) Kagomé-net of Fe atoms shown along the crystallographic c-axis ([0001]); d) Laves phase triple layer as shown in b) but viewed along the c-axis with the slip directions indicated for the synchroshear mechanism: 1. Slip of Fe-layer c (grey arrow), Mo-layer β and upper crystal (red arrow) by vector $b_2$; 2.Synchronous slip of Mo-layer β and upper crystal by vector $b_3$ into free spaces left behind by Fe c-layer atoms to avoid Mo-Mo atom collision in Mo-layer α. The resulting Burgers vector –$b_1$ of this operation is shown as dotted line.

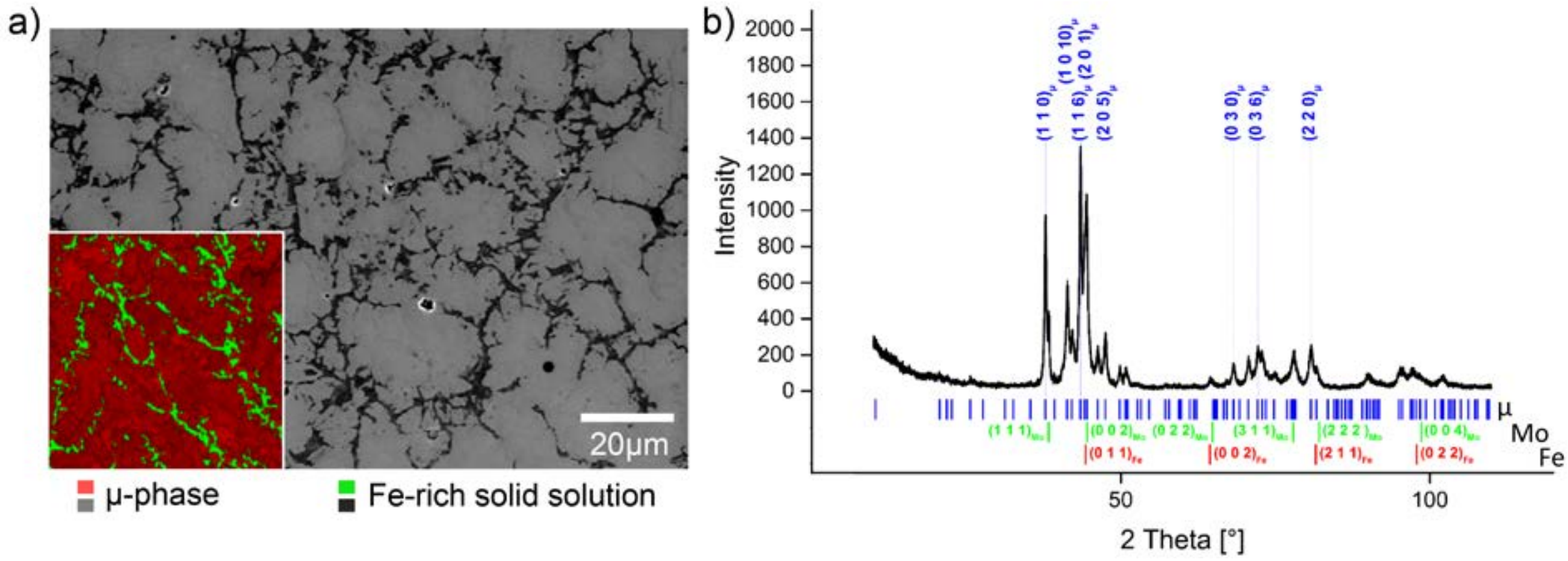

Figure 2: (a) Secondary electron micrograph, (b) X-ray diffractogram, in the XRD spectrum, the main Miller indices of the constituent phases are explicitly shown, additionally, the positions of the diffraction peaks of the µ-phase, α-Fe and α-Mo, are given below the XRD spectrum: blue lines represent the positions of µ-phase peaks, red lines mark the positions of Fe peaks and green lines the positions of Mo peaks.

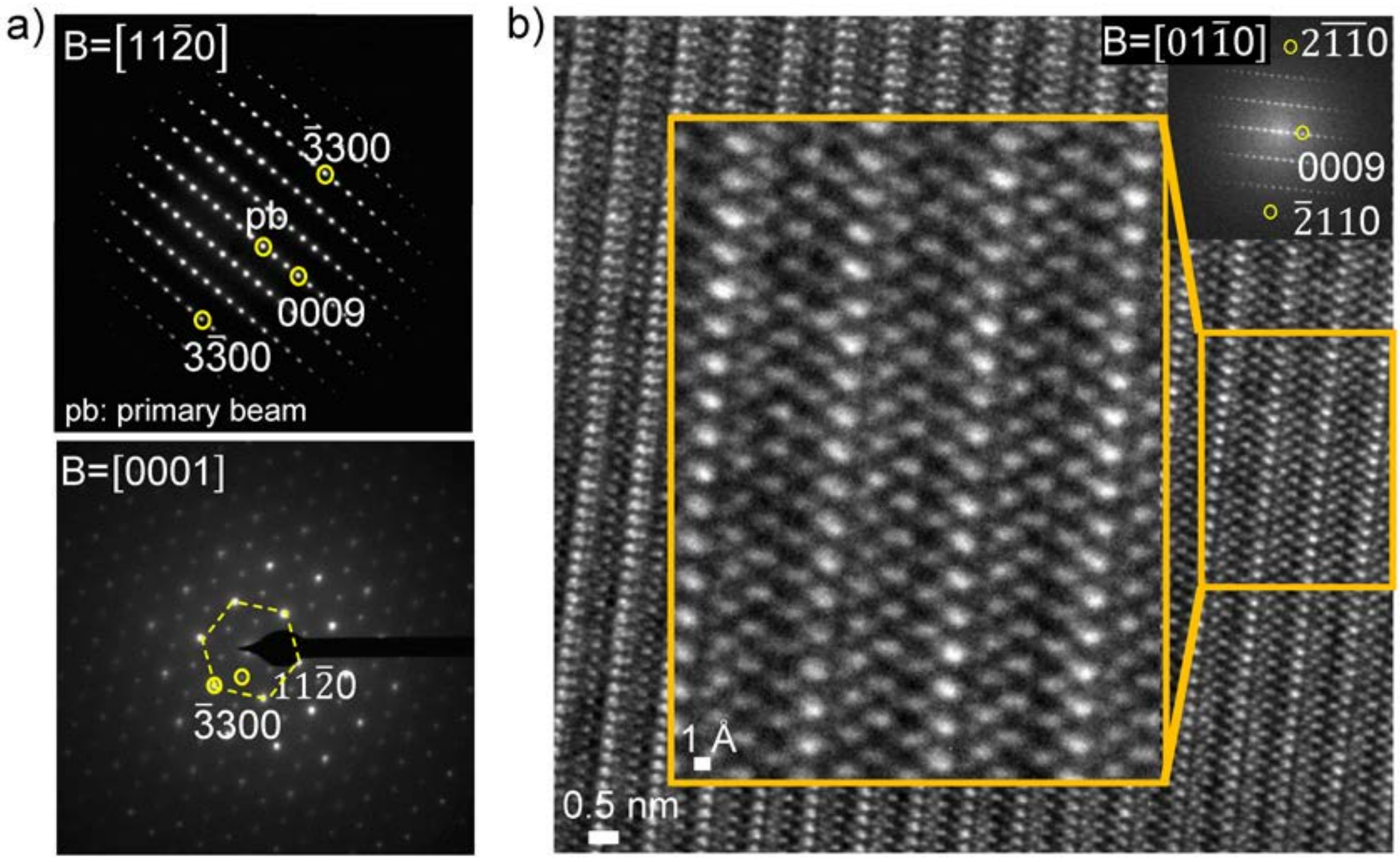

Figure 3: (a) selected area electron beam diffraction (SAED) pattern of the $[11\bar{2}0]$ and $[0001]$ zone axes, and (b) HR-TEM micrograph with $[01\bar{1}0]$ zone axis orientation of the heat-treated Fe-Mo 55 wt.-% alloy.

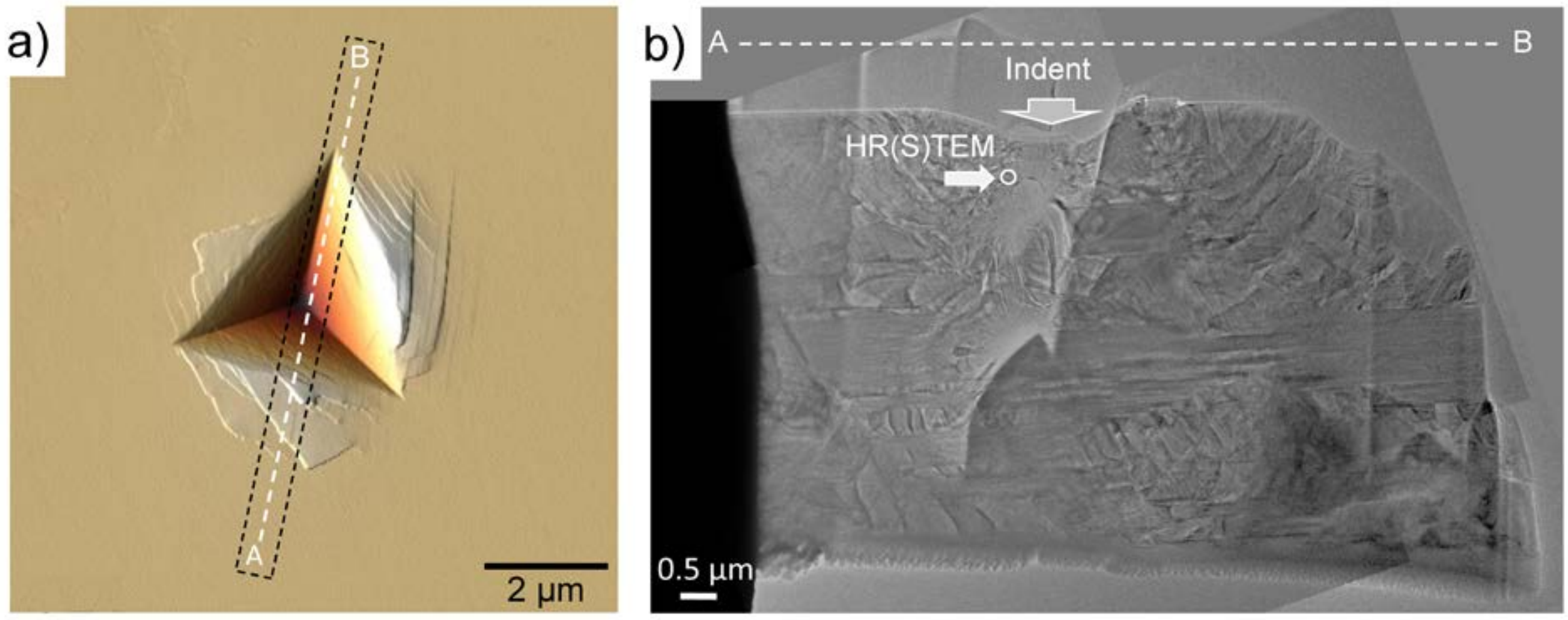

Figure 4: (a) Atomic force microscope (AFM) image of the indent. The dashed rectangle highlights the position of the TEM lamella. (b) Collage of low resolution TEM micrographs of the TEM lamella showing the residual indentation imprint and the area of high resolution TEM investigations presented in Figures 5 and 6. The orientation of the TEM lamella with respect to the indent is indicated by the centre line A-B of the rectangle in the AFM micrograph.

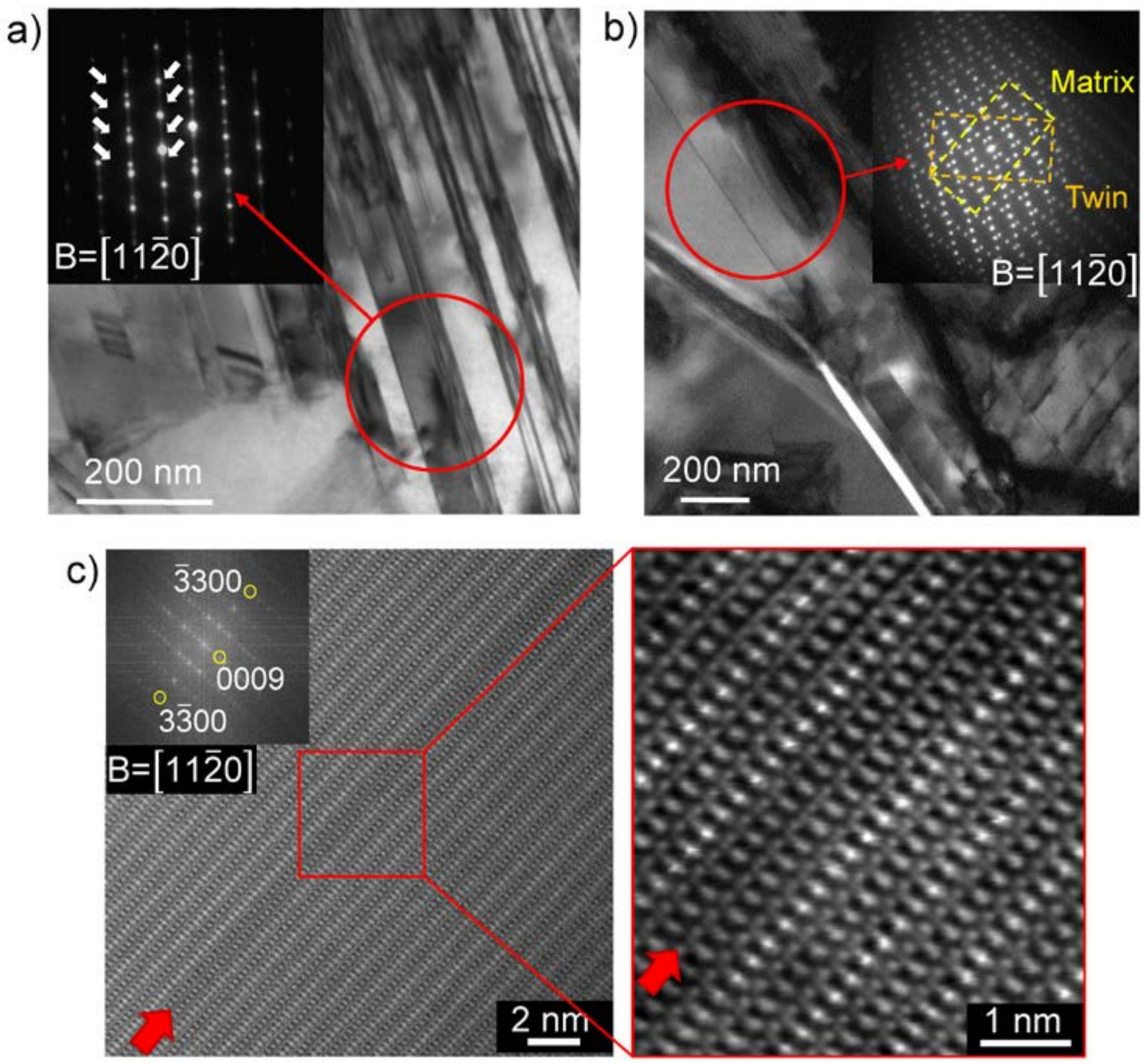

Figure 5: TEM and HAADF-HR-STEM micrographs of grow-in defects in the µ-phase; (a) TEM micrograph and corresponding SAED pattern of a twin-like stacking fault with the $Zr_4Al_3$-layers as mirror planes, (b) TEM micrograph and corresponding SAED pattern of a growth twin, (c) basal stacking fault with disturbed stacking order, here a $Zr_4Al_3$ layer is missing.

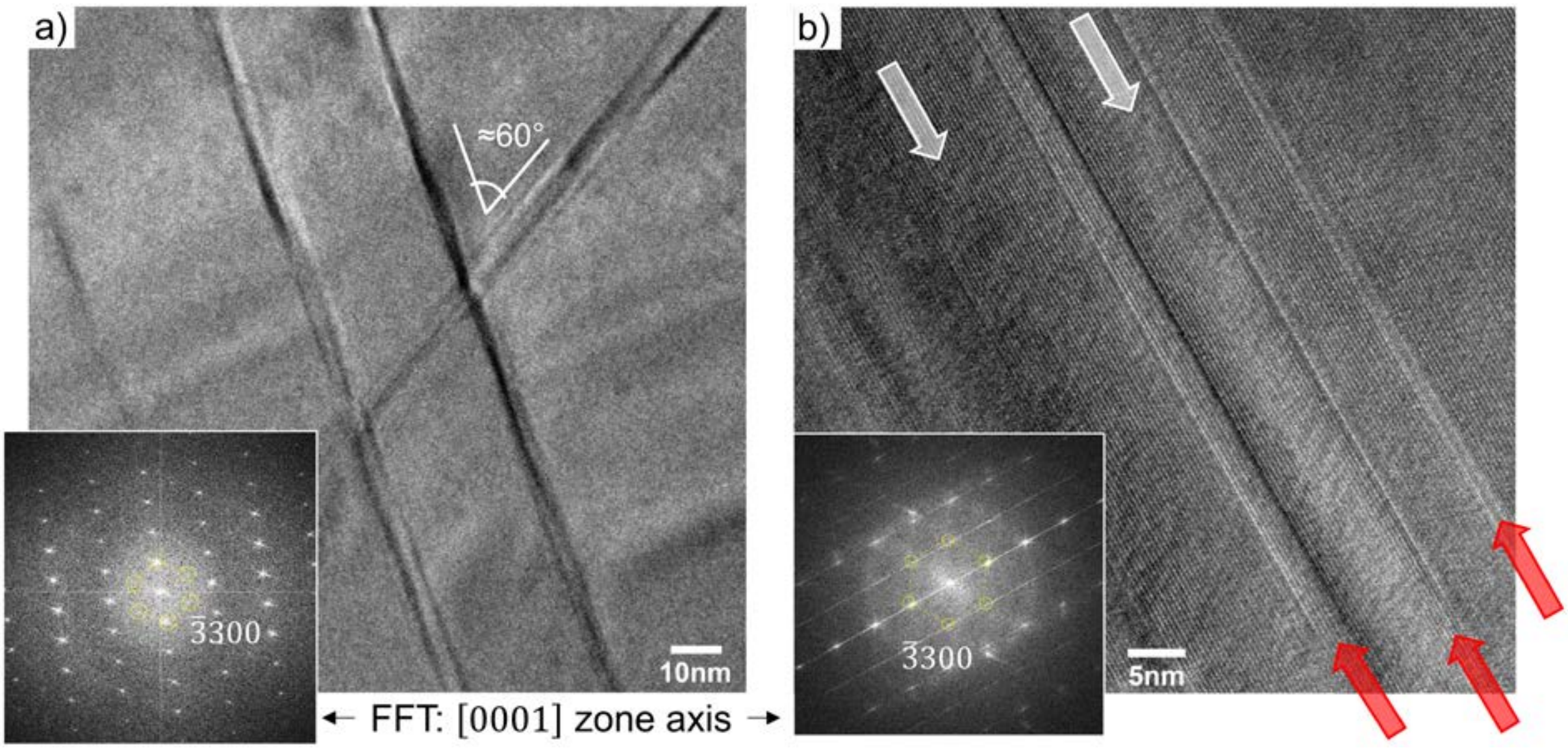

Figure 6: a) HR-TEM image and corresponding FFT pattern of a TEM lamella taken from an area in the plastic deformation zone of a nanoindent showing intersecting stacking faults with an embedded angle of approximately 60°; b) Higher magnification HR-TEM image and corresponding FFT pattern of stacking faults. Red arrows point towards cascades of stacking faults, grey arrows highlight lattice rotation induced by these stacking faults. The corresponding FFT patterns are shown as insets, indicating the incident beam direction to be along the c-axis of the µ-phase.

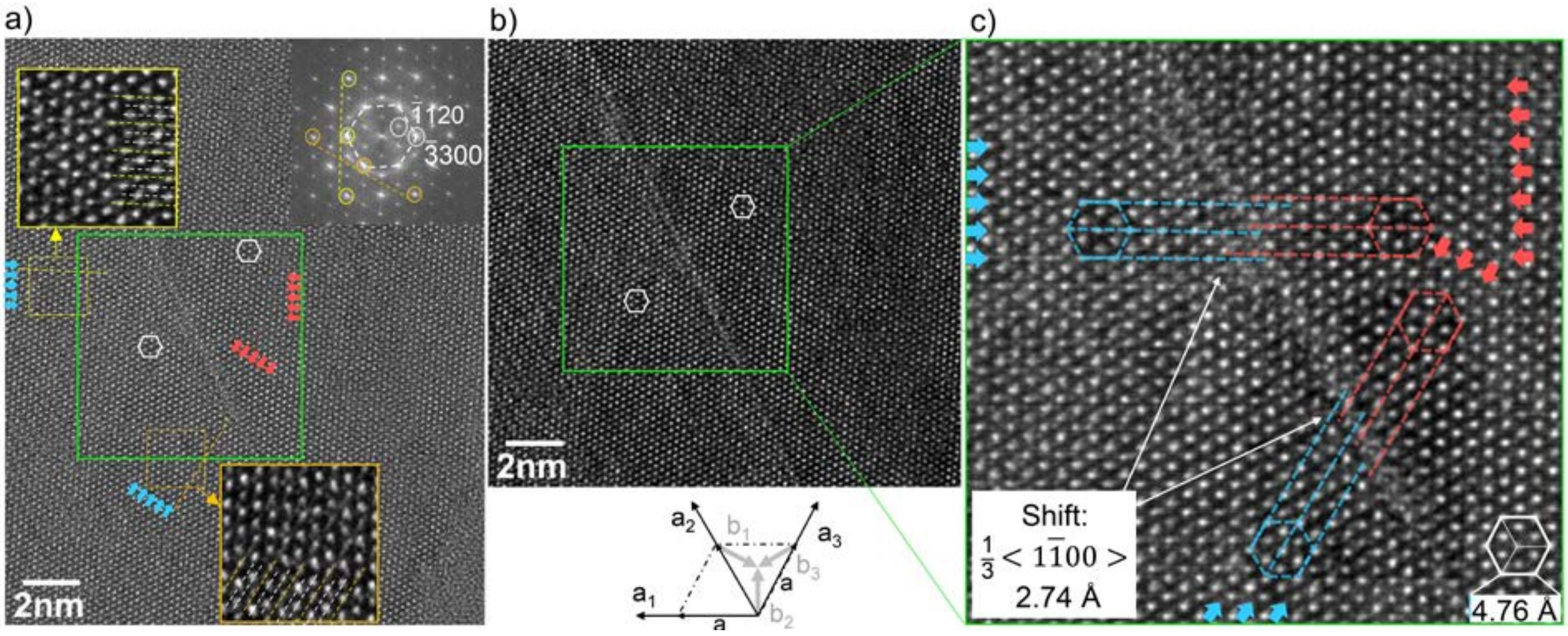

Figure 7: Filtered (left) and unfiltered (middle) high resolution TEM micrographs of a stacking fault below an indent in deformed $Fe_7Mo_6$ µ phase. Bright dots represent the positions of combined Mo/Fe atomic columns. The unit cell of the crystal on both sides of the stacking fault is outlined by white hexagons. The insets in the left micrograph show lines of brighter contrast which correspond to atomic layers with a higher density of Mo atoms, i.e. Laves layer tetrahedrons. The FFT further reveals these Laves layer tetrahedrons along the $\langle 1\bar{1}00 \rangle$ directions by higher brightness of every third spot along these directions. These are highlighted in yellow and orange for $a_1$ and $a_3$, respectively. The close-up on the right side shows the shift of the crystal unit cells across the stacking fault, as highlighted by the coloured hexagons. A shift of $\frac{1}{3}\langle 1\bar{1}00 \rangle$ (2.74 ± 0.01Å) is apparent in both directions, $b_1$ and $b_2$.

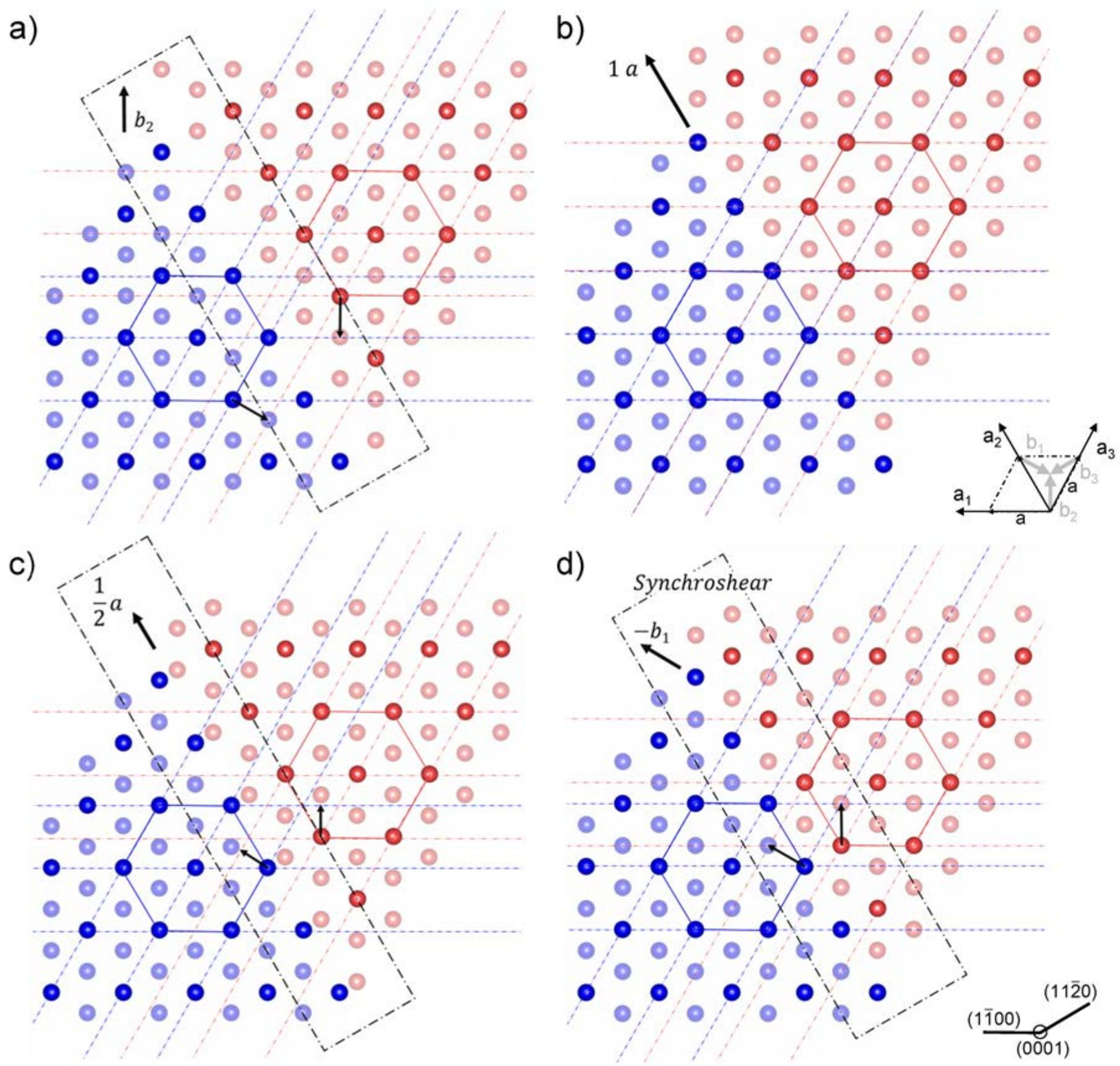

Figure 8: Stacking fault configurations which can be created by the glide of (partial) dislocations in the µ-phase visualised [15] along the crystallographic c-axis. Columns of Mo/ Fe atoms are shown in red and blue, those corresponding to the contour of the unit cell as given in Figure 3 are highlighted. To increase the visibility, columns of only Fe atoms are not shown. The crystal shift induced by (partial) dislocation glide was applied to the upper half of the crystal (coloured in red) and the resulting shift vector is given for each image. a) Partial dislocation slip with $\vec{b}$ = $b_2 = \frac{1}{3}\langle 1\bar{1}00\rangle$; b) Perfect dislocation slip with $\vec{b}$ = a = $\frac{1}{3}\langle 11\bar{2}0\rangle$; c) Shift along 1/2**a** corresponding to slip in stacked-in Kagomé-nets; d) Stacking fault configuration after operation of synchroshear, i.e. glide of synchroshear partials with Burgers vectors $\vec{b}$ = $b_2$ and $\vec{b}$ = $b_3$.

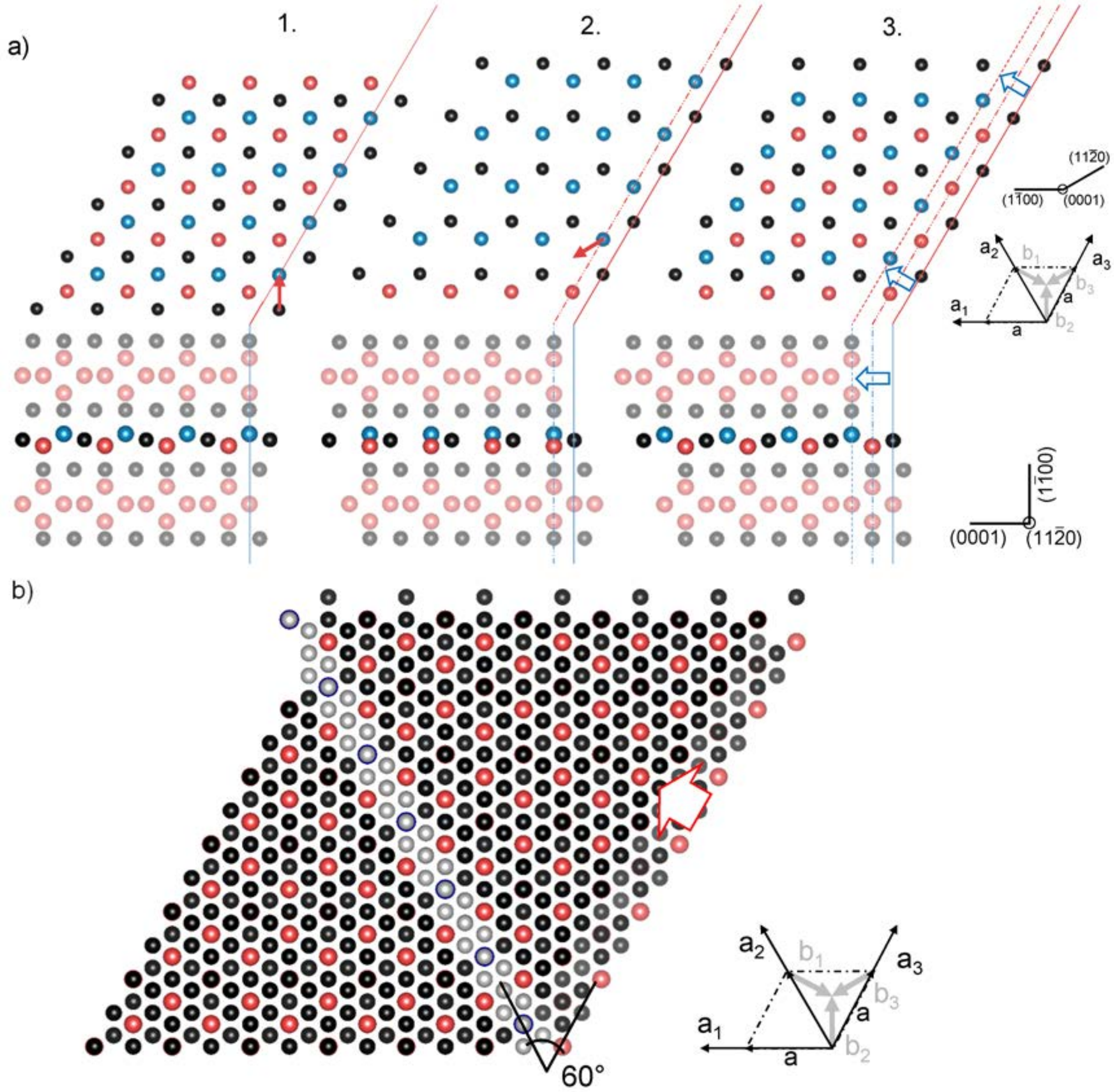

Figure 9: Atomic configurations associated with synchroshear; a) Triple layer of Laves phase illustrating the three stages of the synchroshear mechanism: Undistorted crystal (1.), glide of the first synchroshear partial, $b_2$, shifting the Fe atoms in the Laves layer (black) along $b_2$ (2.) followed by glide of the second synchroshear partial, $b_3$, shifting the Mo atoms (blue) back to the space left by the Fe atoms (3.). A resulting shift vector of $-b_1$ is seen in the upper row of images (3.). The upper row of images is seen perpendicular to the (0001) plane and the lower row of images perpendicular to $(11\bar{2}0)$, along $a_3$; black atoms represent Fe, blue atoms represent the "upper" Mo atoms in the triple layer and red atoms represent "lower" Mo atoms in the triple layer, these are considered as "rigid" during synchroshear [43, 44]. b) stacking fault configuration after synchroshear viewed along the crystallographic c-axis. Fe atoms in the stacking fault are highlighted in grey, Mo atoms of the faulted area are shown in blue. Shift was applied to the upper part of the crystal, the resulting vector is $-b_1$ (see arrow). Atoms of the crystal below the stacking fault are shown in grey to mark the crystal shift. At the lower egde of the image the angle between the stacking fault and the reversed triple layer (t') [45] amounts to 60°.